# Experimental realization of epsilon-near-zero metamaterial slabs with metal-dielectric multilayers

Jie Gao[1,†], Lei Sun[1], Huixu Deng[1], Cherian J. Mathai[2], Shubhra Gangopadhyay[2], and Xiaodong Yang[1,‡]

*[1]Department of Mechanical and Aerospace Engineering, Missouri University of Science and Technology, Rolla, MO 65409, USA*
*[2]Department of Electrical and Computer Engineering, University of Missouri, Columbia, MO 65211, USA*

**Abstract**

Epsilon-near-zero (ENZ) metamaterial slabs at visible frequencies based on metal-dielectric multilayers are experimentally realized. Transmission, reflection and absorption spectra are measured and used to determine the complex refractive indices and the effective permittivities of the ENZ slabs, which agree with the results obtained from both the numerical simulations and the optical nonlocalities analysis. Furthermore, light propagation in ENZ slabs and directional emission from ENZ prisms are also analyzed. The accurate determination of the ENZ wavelength for metal-dielectric multilayer metamaterial slabs is important for realizing many unique applications, such as phase front manipulation and enhancement of photonic density of states.

Email address: [†]gaojie@mst.edu, [‡]yangxia@mst.edu

Metamaterials are artificially structured composites designed to create extraordinary macroscopic electromagnetic properties that are not achievable in natural materials [1, 2]. Recently, epsilon-near-zero (ENZ) metamaterials emerge into the focus of the extensive exploration due to their anomalous electromagnetic features at microwave and optical frequencies. For instance, with the extremely large wavelength of the electromagnetic wave, ENZ metamaterials can be used to squeeze and tunnel electromagnetic energy through narrow waveguide channels [3-6]. Additionally, the near-zero phase variation of the electromagnetic wave inside ENZ metamaterials suggests that these materials can be used for directive emission and radiation phase front shaping [7, 8]. The inherent negative polarizability of the ENZ metamaterials also enables potential applications in electromagnetic transparency and invisible cloaking [9-12]. Moreover, displacement current insulation in nanocircuits and subwavelength image transporting have also been realized with ENZ metamaterials [13, 14]. The unique ENZ properties can also greatly enhance the optical nonlinearities [15] and photonic density of states [16].

In nature, materials with near-zero permittivity are readily available as noble metals, doped semiconductors [17], polar dielectrics [14], and transparent conducting oxides (TCOs) [18]. The permittivity can be characterized by Drude or Drude-Lorentz model, and the near-zero permittivity is obtained when the frequency approaches to the plasma frequency. Since noble metals exhibit the plasma frequencies in the ultraviolet region, inclusions of noble metals can be embedded in a dielectric host medium to synthesize ENZ materials at visible or near infrared frequencies based on the effective medium theory [19]. ENZ behavior can be realized by utilizing metal coated waveguides at the cutoff frequency [16, 20], silver or gold nanowires array grown in porous alumina template [21, 22], and metal-dielectric multilayer structures [23, 24]. Multilayer metamaterials have been designed to realize unique optical functionalities such as negative refraction [25, 26], hyperlenses for sub-diffraction imaging [27], and indefinite cavities [28]. Here, ENZ metamaterial slabs at visible frequencies based on two types of metal-dielectric multilayers, including Au-$Al_2O_3$ pairs and Au-$SiO_2$ pairs, are fabricated and characterized to study their unique optical properties and locate the ENZ wavelengths. The complex refractive indices and effective permittivities of these ENZ metamaterial slabs are determined from the measured spectral transmittance and reflectance, which match the results obtained from finite element method (FEM) numerical simulations and optical nonlocalities analysis. It is demonstrated that the ENZ wavelengths of metal-dielectric multilayer metamaterial slabs can be accurately obtained from the measured transmission and reflection spectra. In addition, light propagation in ENZ metamaterial slabs and directional emission from an ENZ prism are analyzed to explain the unique ENZ response. The demonstrated ENZ metamaterial slabs at visible frequencies will offer many opportunities in studying intriguing optical behaviors, such as phase front shaping [8], enhanced optical nonlinearities [15], spontaneous emission control [29], and active metamaterials [30].

Fig. 1 shows the schematic of metal-dielectric multilayer structures having uniaxial anisotropic permittivity tensor and unity permeability, with layer permittivities of $\varepsilon_1$, $\varepsilon_2$ and layer thicknesses of $d_1$, $d_2$, where 1 and 2 represent metal and dielectric, respectively. It is known that such multilayer structures have strong optical nonlocal effects [31, 32], where the permittivity components will depend on both frequencies and wave vectors. In order to take into account the optical nonlocalities, multilayer structures can be considered as a one-dimensional photonic crystal along the $x$-direction, which leads to the dispersion relation for eigenmodes with transverse-magnetic (TM) polarization according to the transfer matrix method,

$$\cos(k_x(d_1 + d_2)) = \cos(k_x^{(1)}d_1)\cos(k_x^{(2)}d_2) - \frac{1}{2}\left(\frac{\varepsilon_1 k_x^{(2)}}{\varepsilon_2 k_x^{(1)}} + \frac{\varepsilon_2 k_x^{(1)}}{\varepsilon_1 k_x^{(2)}}\right)\sin(k_x^{(1)}d_1)\sin(k_x^{(2)}d_2) \quad (1)$$

where $k_x^{(1,2)} = \sqrt{k_0^2 \varepsilon_{1,2} - k_y^2}$ are x components of wave vectors in the two layers, and $k_0 = \omega/c$ is the wave vector in free space. Eq. (1) can be written in the form of

$$\frac{k_x^2}{\varepsilon_y^{\text{eff}}} + \frac{k_y^2}{\varepsilon_x^{\text{eff}}} = k_0^2 \quad (2)$$

with the effective permittivity components of $\varepsilon_x^{\text{eff}}(\omega, k)$ and $\varepsilon_y^{\text{eff}}(\omega, k)$ including the nonlocal effects due to the multilayer structures. For TM polarized light at normal incidence with $k_y = 0$, $\varepsilon_y^{\text{eff}} = k_x^2/k_0^2$ will represent the effective dielectric constant of the metamaterial slab and it can be solved from Eq. (1) as,

$$\varepsilon_y^{\text{eff}} = \frac{\arccos^2\left[\cos(\sqrt{\varepsilon_1}k_0d_1)\cos(\sqrt{\varepsilon_2}k_0d_2) - \frac{1}{2}(\sqrt{\varepsilon_1/\varepsilon_2} + \sqrt{\varepsilon_2/\varepsilon_1})\sin(\sqrt{\varepsilon_1}k_0d_1)\sin(\sqrt{\varepsilon_2}k_0d_2)\right]}{k_0^2(d_1+d_2)^2} \quad (3)$$

Two types of metal-dielectric multilayer metamaterial slabs with different material combinations and total thicknesses are fabricated on top of quartz substrates, including gold-alumina (Au-$Al_2O_3$) multilayers (Slab 1) and gold-silica (Au-$SiO_2$) multilayers (Slab 2). Each kind of material is individually deposited on quartz substrate first to calibrate and optimize the deposition parameters for both the electron-beam evaporation system (Kurt J. Lesker) and the magnetron sputtering system (AJA International). The optical constants of materials and the film thicknesses are characterized with the variable angle spectroscopic ellipsometry (VASE, J. A. Woollam Co. VB400/HS-190) by measuring the standard ellipsometric parameters Ψ and Δ at different incident angles. The VASE measurements show that optical constants of Au match the standard data of Johnson and Christy [33], based on the fitting from a general oscillator model. While the dielectric constants of $Al_2O_3$ and $SiO_2$ are fitted from the Cauchy dispersion relation. The VASE measured film thickness for each material also matches the thickness value for the set deposition parameters. Slab 1 is composed of 4 pairs of 18 nm Au layer and 81 nm $Al_2O_3$ layer, where each layer is deposited with the electron-beam evaporation system. Au is deposited at the rate of 0.4 Å/sec and $Al_2O_3$ is deposited at 0.2 Å/sec. Slab 2 is made of 13 pairs of 20 nm Au layer and 80 nm $SiO_2$ layer, deposited with the magnetron sputtering system. Au is deposited at 4 mTorr with 20 sccm of Ar gas flow at 30 W DC power, while $SiO_2$ is deposited at 4 mTorr with 20 sccm of Ar gas flow as well as 5 sccm of $O_2$ gas at 250 W RF power. Both deposition processes can achieve high-quality multilayer metamaterial slabs with low surface roughness. Fig. 2 shows the scanning electron microscope (SEM) pictures of the cross sections of the fabricated metal-dielectric multilayer metamaterial slabs, where the focused ion beam (FIB) system (Helios Nanolab 600) is used to cut the cross sections. Each deposited thin layer can be clearly seen, where the bright and dark stripes correspond to gold layers and dielectric layers, respectively. The thickness of the deposited layers can be characterized with the VASE. For Slab 1, the measured averaged thickness for the Au layer and the $Al_2O_3$ layer is 18 ±0.4 nm and 81 ± 1.6 nm, respectively. For Slab 2, the measured averaged thickness for the Au layer and the $SiO_2$ layer is 20 ±0.5 nm and 80 ±1.5 nm.

In order to characterize the optical properties of fabricated ENZ metamaterial slabs, optical transmission (*T*) and reflection (*R*) spectra at normal incidence are measured with a halogen white light source and a spectrometer. The transmission spectrum is normalized with a quartz wafer, and the reflection spectrum is normalized with a silver mirror. Optical absorption (*A*) spectrum is then obtained from *A* = 1 - *T* - *R*. Fig. 3(a) and Fig. 3(b) plot the measured *T*, *R* and *A* for Slab 1 and Slab 2 respectively, together with the FEM simulation results using the VASE measured layer thicknesses and optical constants of materials. Fig. 3(c) and Fig. 3(d) show the complex refractive indices ($n + ik$) for both the metamaterial slabs determined from the measured transmission $T$ and reflection $R$ only, using an approximated approach based on incoherent interference of optical beam oscillated inside a slab at normal incidence (Method 1) [34, 35]. It is emphasized that this approach is practical since it does not require the phase information for the transmission and reflection data. The following relations are used to retrieve the values of $n$ and $k$ of the metamaterial slabs [35]:

$$k = \frac{-\lambda}{4\pi t} \ln \left\{ \frac{[T^2-(1-R)^2]+\{[T^2-(1-R)^2]^2+4T^2\}^{1/2}]}{2T} \right\} \quad (4)$$

$$n = \frac{(1+R_{as})}{(1-R_{as})} + \left[\frac{4R_{as}}{(1-R_{as})^2} - k^2\right]^{1/2} \quad (5)$$

where the surface reflectance between air and metamaterial slab at normal incidence $R_{as}$ is

$$R_{as} = \frac{R}{1+\left[[T^2-(1-R)^2]+\{[T^2-(1-R)^2]^2+4T^2\}^{1/2}\right]/2} \quad (6)$$

The complex refractive indices ($n + ik$) are also retrieved from the FEM simulated $S$ parameters of reflectance $S_{11}$ and transmittance $S_{21}$ (including both the amplitude and phase), according to the algorithm in Ref. [36] (Method 2). In addition, the values of $n$ and $k$ calculated from the nonlocal dispersion relation based on Eq. (3) are plotted for comparison (Method 3). Note that the effective permittivity $\varepsilon_y^{\text{eff}} = (n + ik)^2$ since the permeability $\mu = 1$ for the multilayer slabs. As shown in Fig. 3(c) and Fig. 3(d), the complex refractive indices calculated from Method 2 and Method 3 agree with each other. The refractive indices determined from the measured transmission and reflection without the phase information in Method 1 reproduce the same trend as the calculation results from Method 2 and Method 3. Especially, the ENZ wavelength where $n$ is equal to $k$ overlaps and thus $\text{Re}(\varepsilon_y^{\text{eff}}) = 0$ is located at 659.764 nm for Slab 1 and 620.347 nm for Slab 2. Fig. 3(e) and Fig. 3(f) give the effective permittivity obtained from the relation of $\varepsilon_y^{\text{eff}} = (n + ik)^2$, according to the above three methods. The location of ENZ wavelength where $\text{Re}(\varepsilon_y^{\text{eff}}) = 0$ is predicted accurately with Method 1, although the slope of $\text{Re}(\varepsilon_y^{\text{eff}})$ is slightly different from the calculated values from Method 2 and Method 3. The $\text{Im}(\varepsilon_y^{\text{eff}})$ shows a good trend compared to the calculation results. Based on the above analysis, the ENZ wavelength can be determined precisely for metal-dielectric multilayer metamaterial slabs according to the measured transmission and reflection spectra only, which is critically important for realizing applications based on ENZ metamaterials.

At the ENZ wavelength of 620.347 nm for Slab 2, the calculated effective permittivity is $\varepsilon_y^{\text{eff}} = 0.0 + 0.213i$ and the complex refractive index is $n + ik = 0.326 + 0.326i$, based on the optical nonlocality analysis. Fig. 4(a) shows the FEM calculated electric field $E_y$ distribution along the propagation direction in the slab at the ENZ wavelength, while Fig. 4(b) plots the corresponding phase for the electric field $E_y$. The electric field amplitude decays inside the ENZ slab due to the optical loss so that the transmission is quite low at the output. The phase of the electromagnetic wave varies much more slowly within the ENZ slab due to the small wave vector, compared to the fast phase change outside the slab. Functional optical devices can be constructed by the ENZ metamaterial slabs. Fig. 5(a) shows a schematic of a prism made of the ENZ slab in air, with a 10º titled upper surface for realizing the directional emission. The incoming light with vertical wave vector $k_{\text{in}}$ and Poynting vector $S_{\text{in}}$ propagates in the prism as it crosses the lower surface from air, while the refracted out-going light with the wave vector $k_{\text{out}}$ and Poynting vector $S_{\text{out}}$ will be normal to the upper surface. The dispersion relation of metal-dielectric multilayers described by Eq. (2) can be plotted as iso-frequency contours (IFCs) in $k$ space at specified frequencies. Fig. 5(b) shows the IFCs for both the real part and the imaginary part of $k_x$ as a function of $k_y$ with real values at the ENZ wavelength of 620.347 nm for Slab 2, when the TM-polarized light propagates inside. Physically, the directional emission is resulted in the conservation of the tangential component of the wave vectors at the upper interface between the prism and air, which can be explained by the flat elliptical IFC of the ENZ metamaterial compared with the circular IFC of air, as illustrated in Fig. 5(b). It is clear that the propagating direction of the out-going light will be very close to the normal direction of the upper interface of the prism even though the material loss is considered. Fig. 5(c) gives the FEM simulations of the directional emission for prisms made of Au-SiO$_2$ multilayers with a 10$^o$ tilted surface at the ENZ wavelength. The distribution of the electric field amplitude $E_y$ is plotted, with the power flow indicated by hollowed arrows. Since $\varepsilon_y^{\text{eff}}$ is near zero, according to Maxwell's equation, the electric field $E_y$ in the ENZ slab satisfies the static-like equation, indicating that the phase velocity is infinitely large and the phase variation is very small. Light will be refracted at the prism upper surface with a small angle of refraction and the phase front will be conformal to the surface shape due to the near-zero permittivity $\varepsilon_y^{\text{eff}}$.

In this work, metal-dielectric multilayer ENZ metamaterial slabs have been experimentally realized at visible frequency region. The thickness and optical constants of each deposited layer in multilayer metamaterial slabs are characterized and the transmission and reflection spectra are measured to understand their optical properties. The complex refractive indices and the effective permittivities of the multilayer metamaterial slabs are determined from the measured transmission and reflection, which match the retrieved values from numerical simulations and the results obtained from optical nonlocalities analysis. Moreover, light propagation in ENZ metamaterial slabs and directional emission from an ENZ prism are simulated to explain the unique optical response at the ENZ wavelength. The demonstrated ENZ metamaterial slabs will enable many promising applications, such as phase front shaping, enhanced optical nonlinearities, photonic density of states manipulation, and active optical metamaterials.

**Acknowledgments**
This work was partially supported by the Intelligent Systems Center and the Materials Research Center at Missouri S&T, the University of Missouri Interdisciplinary Intercampus Research Program, the University of Missouri Research Board, and the Ralph E. Powe Junior Faculty Enhancement Award.

**Figures**

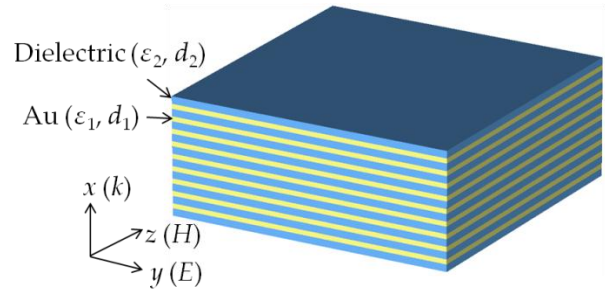

Fig. 1. Schematic of metal-dielectric multilayer metamaterial slabs, with layer permittivities of $\varepsilon_1$, $\varepsilon_2$ and layer thicknesses of $d_1$, $d_2$, where 1 and 2 represent metal and dielectric respectively. The incident light is TM polarized along the $x$ direction with the components of $E_y$ and $H_z$.

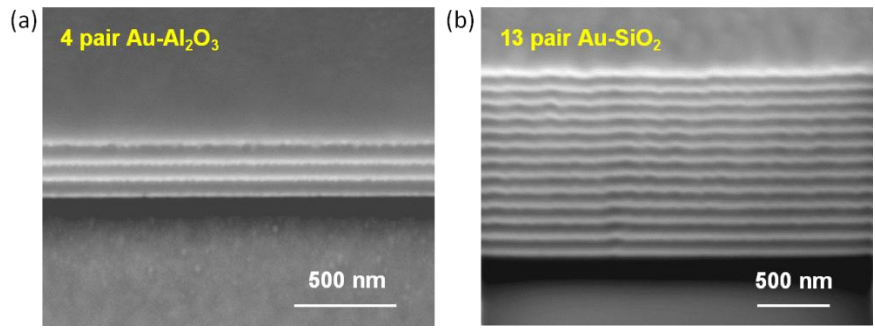

Fig. 2. SEM pictures of the cross sections of fabricated metal-dielectric multilayer metamaterial slabs made of **(a)** 4 pairs of 18 nm Au and 81 nm $Al_2O_3$ multilayers (Slab 1) and **(b)** 13 pairs of 20 nm Au and 80 nm $SiO_2$ multilayers (Slab 2). The bright and dark stripes correspond to gold and dielectric layers, respectively. The layer thickness is characterized with the VASE.

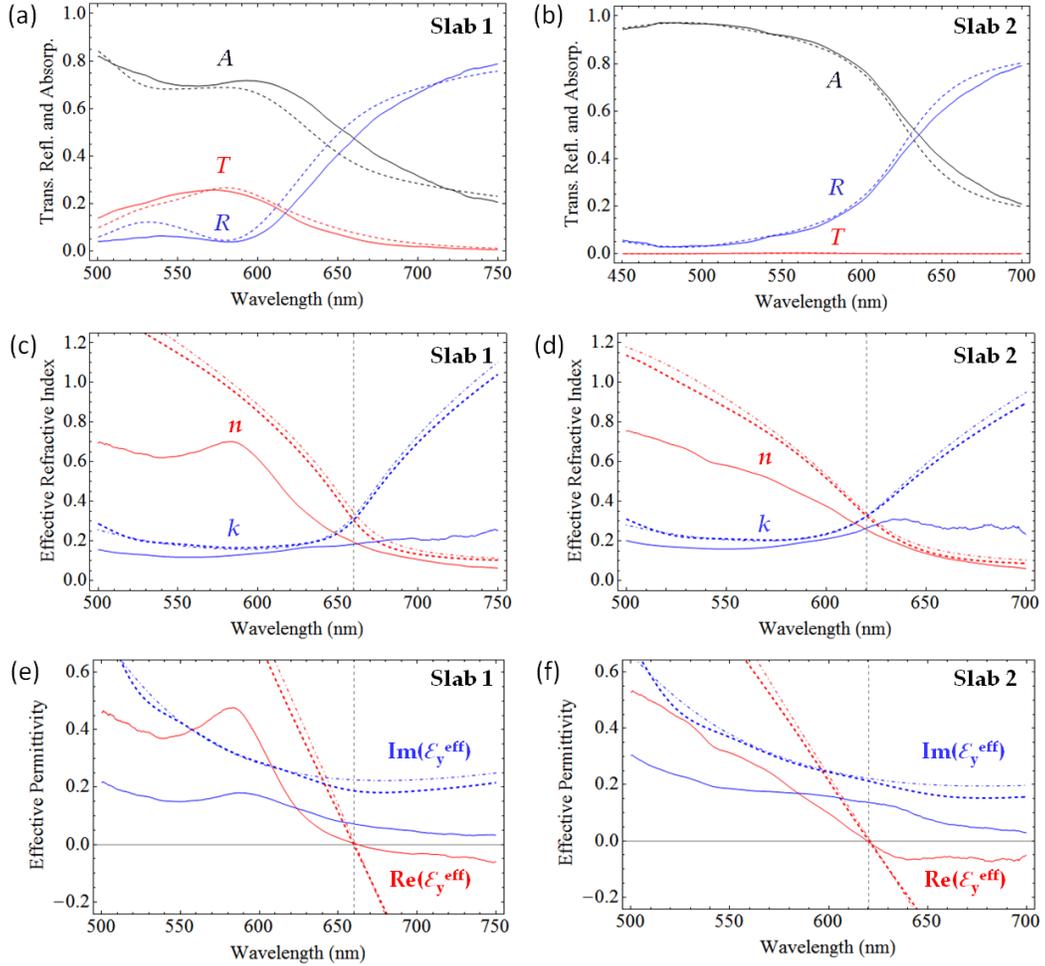

Fig. 3. **(a), (b)** Transmission ($T$), reflection ($R$), and absorption ($A$) spectra for the 4 pairs of Au-$Al_2O_3$ multilayers (Slab 1) and the 13 pairs of Au-SiO2 multilayers (Slab 2). The solid curves represent the measured data, while the dashed curves show the FEM simulation results. **(c), (d)** The retrieved complex refractive indices $n$ and $k$ for each metamaterial slab. The solid curves represent the values determined from the measured transmission and reflection (Method 1). The dashed curves represent the FEM simulation retrieved values (Method 2), and the dotted curves show the results of optical nonlocality analysis (Method 3). **(e), (f)** Real and imaginary parts of the effective permittivity $\varepsilon_y^{eff}$ obtained from the three methods.

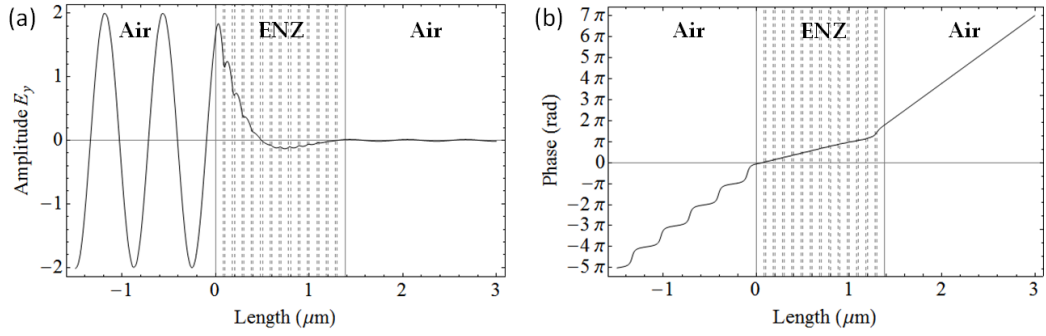

Fig. 4. **(a)** The electric field $E_y$ distribution and **(b)** the phase profile of electric field along the propagation direction for Slab 2 at the ENZ wavelength of 620.347 nm. The decay of the electric field amplitude is due to the optical loss. The phase varies slowly within the ENZ slab compared to the fast phase change in air. The dashed line represents the interface of Au layer and $SiO_2$ layer.

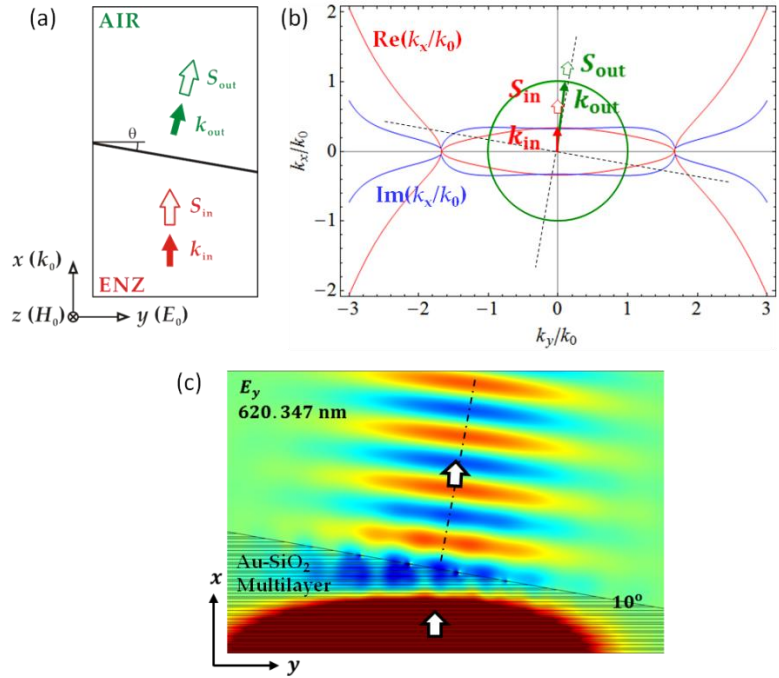

Fig. 5. **(a)** The schematic diagram of a prism made of the ENZ slab in air for the directional emission at the ENZ wavelength. **(b)** The IFCs at the ENZ wavelength for the 13 pair Au-$SiO_2$ multilayer structures, where both the real part (red solid curve) and the imaginary part (blue solid curve) of $k_x$ as a function of $k_y$ with real values are plotted. The IFC of air is plotted as a green circle. **(c)** FEM simulated directional emission of the ENZ prism made of Au-$SiO_2$ multilayers with a $10°$ titled upper interface, where the electric field $E_y$ is shown.